\documentclass[
superscriptaddress,
twocolumn,
amsmath,amssymb,
aps,
prl,
]{revtex4-2}

\usepackage{graphicx,framed} 
\usepackage{dcolumn} 
\usepackage{bm} 
\usepackage{braket}
\usepackage{dsfont}
\usepackage{color}
\usepackage{amsthm}
\usepackage{hyperref} 
\usepackage[normalem]{ulem} 
\usepackage{qcircuit}       

\usepackage{amsthm}

\usepackage{comment}

\usepackage{algorithm}
\usepackage[noend]{algpseudocode}
\algdef{SE}[DOWHILE]{Do}{doWhile}{\algorithmicdo}[1]{\algorithmicwhile\ #1}
\makeatletter
\def\algbackskip{\hskip-\ALG@thistlm}
\makeatother

\definecolor{lightblue}{RGB}{73,151,208}
\definecolor{crimson}{RGB}{140,41,53}

\hypersetup{
    colorlinks,
    linkcolor={crimson},
    citecolor={lightblue},
    urlcolor={lightblue}
}

\theoremstyle{definition}

\newcommand{\tr}{\mathrm{Tr}}

\newcommand{\mD}{\mathcal{D}}

\newcommand{\mU}{\mathcal{U}}

\newcommand{\lsec}[1]{\textit{#1.---}}

\begin{document}

\preprint{}

\title{Single-Particle Universality of the Many-Body Spectral Form Factor}

\author{Michael O. Flynn}
\email{moflynn@bu.edu}
\affiliation{Department of Physics, Boston University, Boston, Massachusetts 02215, USA}
\affiliation{Department of Physics \& Astronomy, University of Victoria, Victoria, British Columbia V8P 5C2, Canada}

\author{Lev Vidmar}
\email{lev.vidmar@ijs.si}
\affiliation{Department of Theoretical Physics, J. Stefan Institute, SI-1000 Ljubljana, Slovenia}
\affiliation{Department of Physics, Faculty of Mathematics and Physics, University of Ljubljana, SI-1000 Ljubljana, Slovenia}

\author{ Tatsuhiko N. Ikeda}
\email{tatsuhiko.ikeda@riken.jp}
\affiliation{Department of Physics, Boston University, Boston, Massachusetts 02215, USA}
\affiliation{RIKEN Center for Quantum Computing, Wako, Saitama 351-0198, Japan}

\begin{abstract}
We consider systems of fermions evolved by non-interacting unitary circuits with correlated on-site potentials. When these potentials are drawn from the eigenvalue distribution of a circular random matrix ensemble, the single-particle sector exhibits chaotic dynamics. We study the corresponding many-body spectral statistics and show that the spectral form factor (SFF) can be computed \textit{exactly}. Due to the absence of interactions the SFF grows exponentially in time, a result which we demonstrate through simple arguments, scaling collapses, and closed-form evaluation of the SFF. We study the role of interactions by numerically analyzing a kicked Ising model and find that the SFF crosses over to a linear growth regime consistent with many-body random matrix universality. Our exact results for the SFF provide a baseline for future studies of the crossover between single-particle and many-body random matrix behavior.
\end{abstract}

\maketitle

Recent advances in experimental physics have ushered in an era of unprecedented control over isolated quantum mechanical systems~\cite{BlochQuantumSimulators,QuantumSimulatorReview,RMP_Simulators,Science_Simulators}. These developments have complemented and renewed theoretical interest in fundamental dynamical concepts such as chaos and ergodicity~\cite{ThermalizationOverview,AdiabaticEigenstateDeformations,DefiningClassicalQuantumChaos,OperatorGrowthHypothesis,DeutschStatMech, Srednicki1,Srednicki2, Srenicki3,rigol_dunjko_08, ETHReview}. Famously, it is difficult to even rigorously define these terms and researchers often understand them through the lens of random matrix theory (RMT) and spectral statistics~\cite{Wigner1,Wigner2,Wigner3,RMTreviewOne,RMTReview2,sierant_lewenstein_24}. This perspective on many-body dynamics has proven extremely fruitful, generating active research in high-energy physics, quantum information theory, and condensed matter theory~\cite{RMTBlock1,RMTBlock2,RMTBlock3}.

Historically, the insights derived from RMT have proven particularly successful in analyzing several classes of dynamical systems, including integrable models, chaotic single-particle systems (such as billiards), and the late time dynamics of chaotic many-body systems~\cite{Billiards1,Billiards2, Billiard3,Billiard4}. More recently, new regimes such as systems governed by hydrodynamics have been studied from the perspective of spectral statistics, particularly as they are encoded in the spectral form factor (SFF)~\cite{Kos18, Bertini2018, ChanMinimalModel, Chan2018, Liu2018, ChanConservedCharge, suntajs_bonca_20a, sierant_delande_20, suntajs_prosen_21, Vasilyev20, Prakash21, Dibyendu1, Dibyendu2, Liao2, ErgodicityBreakingZeroDimensions, Joshi22, WinerHydroSFF, Liu17, Gharibyan18,  DissipativeSFF, BrokenUnitaritySFF}. It remains an important problem in many-body dynamics to search for models which are amenable to exact solutions while exhibiting universal dynamical properties.

In this Letter, we analyze models of non-interacting fermions and argue that their \textit{many-body} statistics can be studied exactly. The exactness of our results makes these models strong candidates to analyze the onset of many-body random matrix universality in the presence of interactions; rigorously analyzing the non-interacting regime is an essential first step in this direction. To that end, we compute the SFFs of our fermion models using exact numerical and analytical approaches. For circuit parameters drawn from the circular unitary ensemble (CUE), we report a simple analytic form for the SFF at integer times which consists of a sequence of exponential ramps. After discussing this result and its consequences, we conjecture and numerically verify scaling ansatz for the circular orthogonal and symplectic ensembles (COE and CSE, respectively). Finally, we argue that our results can be applied to interacting Hamiltonian and Floquet systems, including a kicked Ising model which we analyze numerically. Our results hence open new possibilities for detecting single-particle chaos in many-body spectra and dynamics.

\lsec{Model}
Our model draws inspiration from Refs.~\cite{Winer2020,Liao2020}, which studied the many-body spectral statistics of chaotic single-particle Hamiltonians. Instead, we consider unitary circuits with parameters taken from a circular ensemble; for concreteness, we focus here on the CUE. A member of the circuit ensemble is constructed by first drawing an $L\times L$ unitary, $U$, from the CUE and obtaining its eigenvalues, written as $\lambda_{j}=e^{i\theta_{j}}$ ($j=1,2,\dots,L$), where $\theta_j\in\left[-\pi,\pi\right)$. The single-particle quasienergies, $\theta_j$, then define the many-body unitary
\begin{align}\label{eq:Udef}
    \mU = \exp\left( -i\sum_{i=1}^L \theta_i n_i \right),
\end{align}
where $n_i$ is a fermion number operator. The many-body quasienergies of $\mU$ are determined by the fermion occupations, $\Theta(\bm{n}) = \bm{n}\cdot\bm{\theta}$.
We are interested in the spectral statistics of this random matrix ensemble, particularly as they are encoded in the SFF~\cite{BerrySpectralRigidity}
\begin{equation}\label{eq:Kdef}
K_{\mU}(L,t) \equiv \left\langle \left| \tr \left(\mU^t\right) \right|^2  \right\rangle_{\mU},
\end{equation}
where the trace is taken over the many-body Hilbert space and $\langle\cdot\rangle_{\mU}$ denotes averaging over $\mU$, that is, over $\theta_j$. In a minor abuse of notation, we have referred to the ensemble as well as a representative member of the ensemble with the same symbol, $\mU$. Throughout this Letter we assume that $t\in\mathbb{Z}$, which guarantees that the SFF is invariant under ``gauge transformations'' $\theta_j\to\theta_j +2\pi m_j,\; m_j\in\mathbb{Z}$. However, our methods for computing the SFF can be extended to arbitrary real times as discussed in Ref.~\cite{OurPaper2}. 

\begin{figure}
    \centering
    \includegraphics[width=\columnwidth]{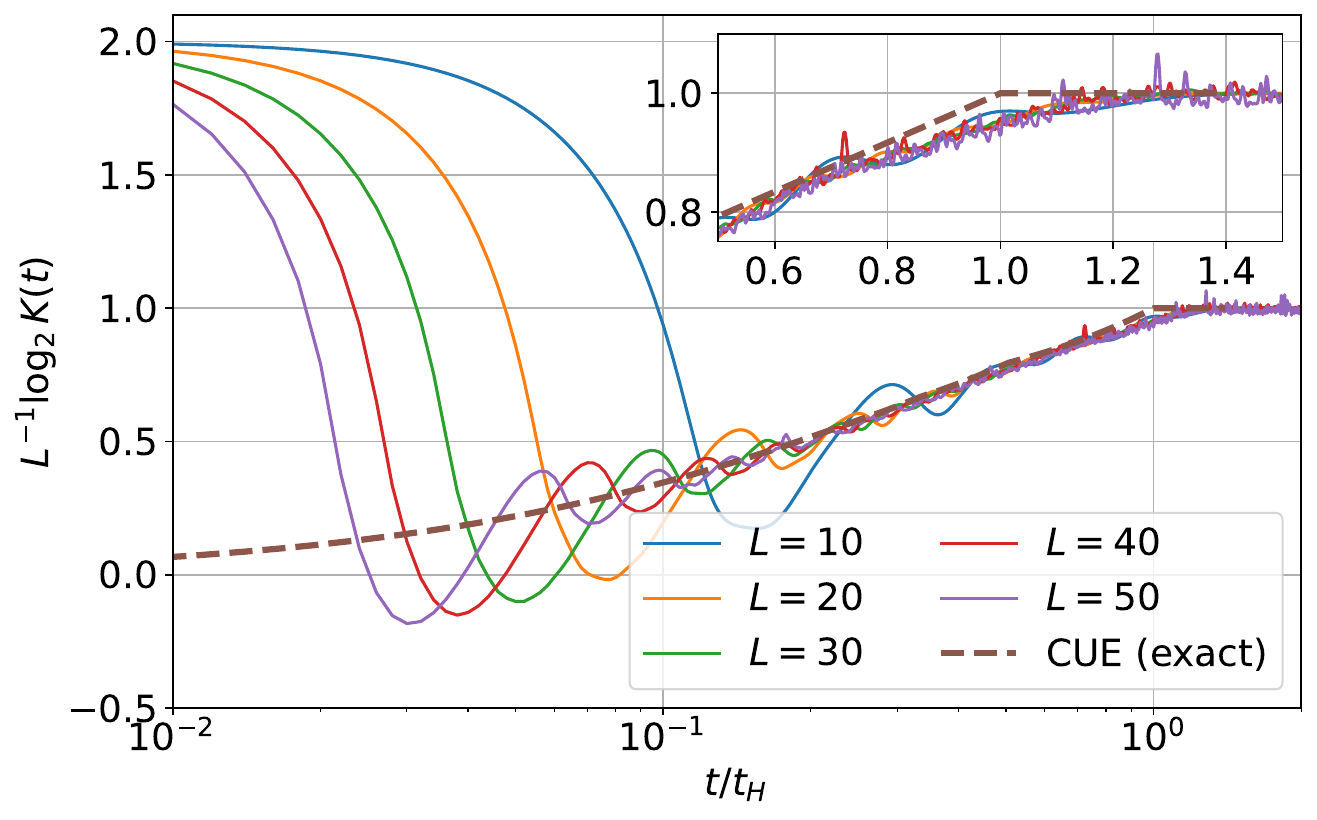}
    \caption{
    Comparison of the many-body CUE and GUE SFFs. The GUE data (solid curves) is averaged over $10^7$ disorder realizations and the CUE prediction is taken from~\eqref{eq:ExactSFF}. The time axis is normalized by the single-particle Heisenberg scales, $t_H^{\text{1, (GUE)}}=\frac{\pi}{\sqrt{2}}L$ and $t_H^{\text{1, (CUE)}}=L$.
    Inset: a magnified view for $t/t_H^{(1)} \geq 1/2$ with linear scales.
    }
    \label{fig:Kt_GUE}
\end{figure}

Several remarks concerning this model are in order. One might guess that the ensemble-averaged properties of $\mU$ are consistent with integrability since the circuits are non-interacting; however, this is incorrect. To see this, note that the single-fermion sector is characterized by the random matrix statistics of the CUE. In particular, the single-particle quasienergies have the joint distribution~\cite{DysonEigenvalueCorrelations,mehta1991random}
\begin{equation}\label{eq:CUEDist}
P(\theta_{1},\cdots,\theta_{L})\propto \prod_{j<k}|e^{i\theta_{j}}-e^{i\theta_{k}}|^{2}.
\end{equation}
The single-particle analog of the SFF in Eq.~\eqref{eq:Kdef}, $K_{U}(L,t)$, is also well-known~\cite{HaakeSignaturesofChaos},
\begin{equation}\label{eq:OneBodySFF}
K_{U}(L,t) = \langle |\tr\left(U^{t}\right)|^{2}\rangle_{U}
=\begin{cases}
        L^{2}\delta(t) + t, & t\leq L\\
        L, & t>L
    \end{cases}
\end{equation}
Both the quasienergy distribution~\eqref{eq:CUEDist} and single-particle SFF~\eqref{eq:OneBodySFF} are inconsistent with Poisson statistics. While neither of these results speak directly to the many-body spectral statistics of $\mU$, we will show that the SFF $K_{\mU}(L,t)$ is also inconsistent with integrability.  Moreover, it is known~\cite{mehta1991random} that the Gaussian unitary ensemble (GUE) and CUE have the same asymptotic level spacing distributions in the large $L$ limit; hence, while the GUE is often used to model Hamiltonian chaos, one could choose to study the CUE in its place while maintaining many features of the GUE (see Fig.~\ref{fig:Kt_GUE} and surrounding discussion).

\begin{figure}
    \centering
    \includegraphics[width=\columnwidth]{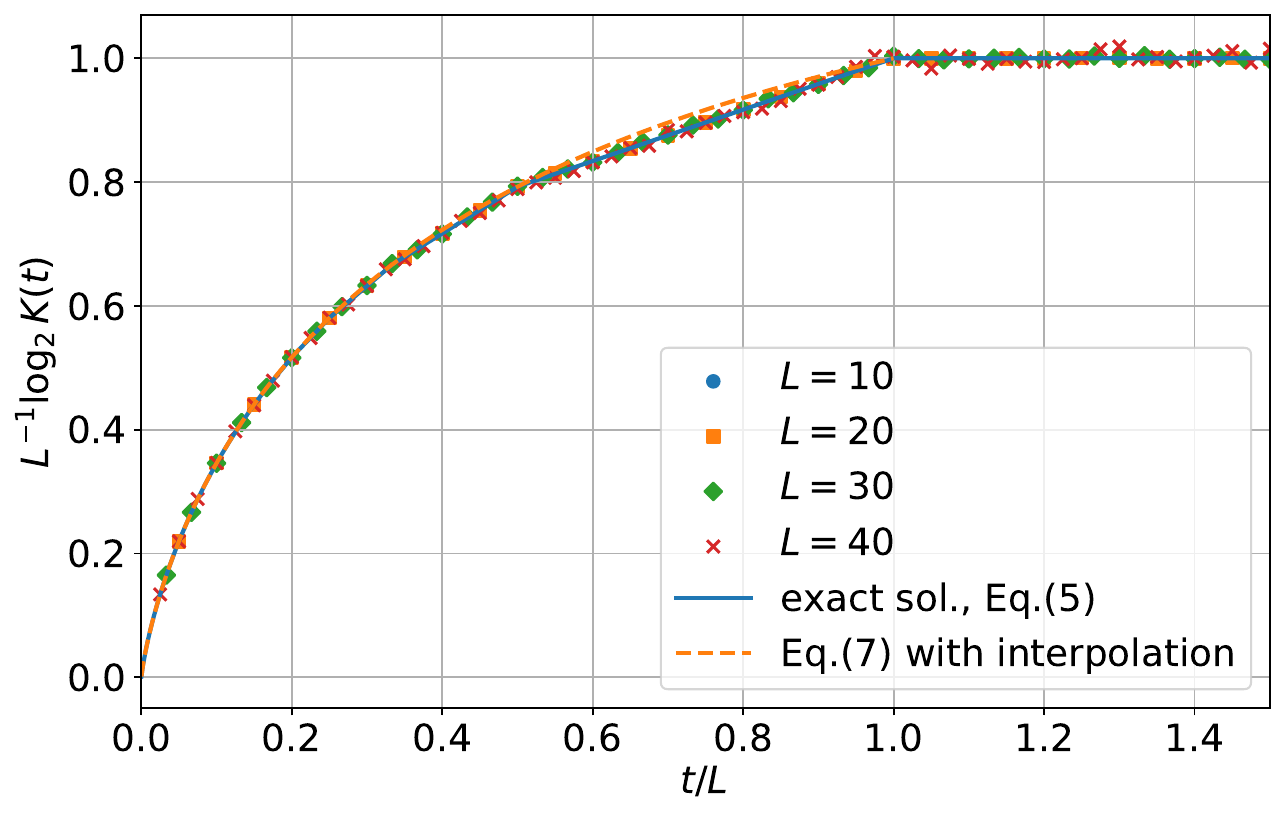}
    \caption{
    Rescaled SFF of the CUE model vs. $t/L$. The finite size data is obtained by averaging Eq.~\eqref{eq:Kdef} over $10^8$ samples. The solid blue curve shows our exact result for the SFF~\eqref{eq:ExactSFF} and exhibits perfect scaling collapse. The orange dashed line shows the smoothened approximation to the exact SFF~\eqref{eq:K_divt}, which is exact when $L/t$ is an integer.
    }
    \label{fig:Kt_demo}
\end{figure}

\lsec{Exact Spectral Form Factor}
We have evaluated the many-body SFF~\eqref{eq:Kdef} in closed form for parameters drawn from the CUE. At $t=0$, the SFF is given by $4^{L}$, while for any integer $t>0$, we set $N=\lfloor L/t\rfloor$ and find
\begin{equation}\label{eq:ExactSFF}
K_{\mU}(L,t>0) = (N+1)^t\left( \frac{N+2}{N+1}\right)^{L-Nt}.
\end{equation}
This result is exact and requires no approximations. Comparisons of~\eqref{eq:ExactSFF} with numerical sampling at fixed $L$ are presented in Fig.~\ref{fig:Kt_demo}. Later in this Letter, we will outline some of the key steps leading to~\eqref{eq:ExactSFF} and present full details in a separate publication~\cite{OurPaper2}. Here, we analyze the consequences of this result.

Before reaching the plateau $K_{\mU}(L,t\geq L)=2^L$, the SFF~\eqref{eq:ExactSFF} consists of a sequence of exponential ramps with growth rates that depend on $L/t$. To see this, fix $L$ and make a list of its divisors: $1=t_{1}<t_{2}<\cdots <t_{M}=L$. Next fix $0<t\leq L$ and find the divisors of $L$ which satisfy $t_{j}\leq t\leq t_{j+1}$. Defining $N_{j} = L/t_{j}\in\mathbb{Z}$, Eq.~\eqref{eq:ExactSFF} yields
\begin{equation}
K_{\mU}(L,t) = K_{\mU}(L,t_j)\exp\left[\lambda_j(t-t_j)\right]\;,
\end{equation}
where $\lambda_j\equiv (N_j+1)\ln (N_j+1) - N_j\ln(N_j+2)$ is the growth rate for $t\in\left[t_{j},t_{j+1}\right)$. In particular, the growth rate jumps when $t$ divides $L$. At late times,$t/L\sim O(1)$, most values of $t$ do not divide $L$ and the SFF is simply described by piecewise exponential functions. When $t/L\ll 1$, the density of divisors of $L$ is large and it is convenient to find an alternative representation for the SFF. When $L/t\in\mathbb{Z}$,
\begin{equation}\label{eq:K_divt}
K_{\mU}(L,t) = \left(\frac{L}{t} +1\right)^{t}\qquad (L/t \in \mathbb{Z}).
\end{equation}
By interpolating this result for times between divisors of $L$, this form qualitatively approximates  the SFF even at late times (see Fig.~\ref{fig:Kt_demo}).

The exponential growth of the SFF~\eqref{eq:ExactSFF} should be contrasted with standard RMT predictions, namely the linear ramp of Eq.~\eqref{eq:OneBodySFF}. The enhancement of the SFF in our model can be understood as a consequence of the relation between single-particle and many-body quasienergies, $\Theta(\bm{n}) = \bm{n}\cdot\bm{\theta}$. This relation, which is absent for interacting systems, implies that dephasing of the $\theta_j$ is sufficient to destroy \textit{all} many-body correlations. These correlations are encoded in the quasienergy distribution~\eqref{eq:CUEDist} and yield an effective (single-particle) Heisenberg time $t_H^{(1)}=L$.

\begin{figure}
    \centering
    \includegraphics[width=\columnwidth]{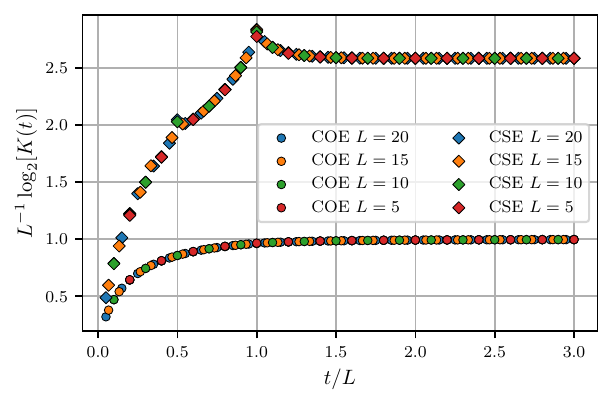}
    \caption{
    The SFF for circuits with parameters drawn from the COE (circles) and CSE (diamonds) over a range of $L$-values. In each case, we have computed the SFF with numerically exact transfer matrix methods which do not require sampling.
    }
    \label{fig:K_COE_CSE}
\end{figure}

\begin{figure}
    \centering
    \includegraphics[width=\columnwidth]{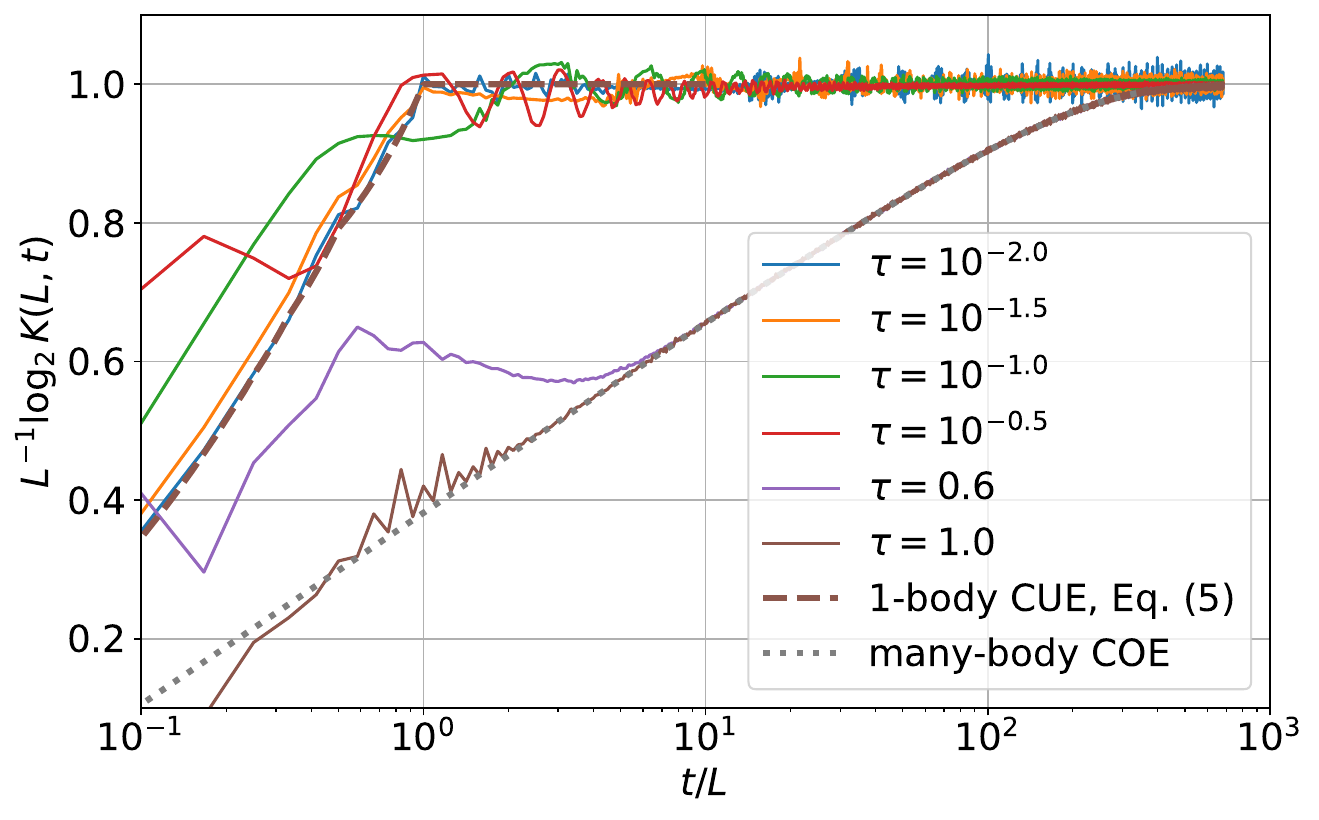}
    \caption{
    The SFF for the kicked Ising model with $L=12$ and various interaction strengths $\tau$, averaged over $10^4$ disorder realizations.
    The dashed (dotted) curve shows the exact solution for the one-body CUE~\eqref{eq:ExactSFF} (the many-body COE~\eqref{eq:K_MB_COE}).
    }
    \label{fig:KIM_SFF}
\end{figure}

The SFF~\eqref{eq:ExactSFF} exhibits scaling in units of the single-particle Heisenberg time. Indeed, the preceding argument suggests that $\log K_\mU(L,t) = Lf(t/L)$ for an undetermined function $f$. This is consistent with the exact CUE result~\eqref{eq:ExactSFF},
\begin{equation}\label{eq:K_in_log}
    \frac{\log_2 K_{\mU}}{L} = \frac{t}{L}\log_2(N+1)+\left(1-\frac{Nt}{L}\right)\log_2 \left( \frac{N+2}{N+1}\right).
\end{equation}
This obviously leads to a scaling collapse since our results are exact (see Fig.~\ref{fig:Kt_demo}); we will see nontrivial scaling collapses later (see Figs.~\ref{fig:K_COE_CSE} and~\ref{fig:KIM_SFF}).

\lsec{Calculation Overview}
Here we provide a sketch of the calculations which lead to~\eqref{eq:ExactSFF}. A full presentation of this and related results will be published separately~\cite{OurPaper2}. The SFF~\eqref{eq:Kdef} can be written as
\begin{equation}
K_{\mU}(L,t)=2^{L}\int d\theta_{1}\cdots d\theta_{L}\;P(\bm{\theta})\prod_{j=1}^{L}\left[1+\cos\left(t\theta_{j}\right)\right],
\end{equation}
where $P(\bm{\theta})$ is defined in~\eqref{eq:CUEDist}. Using well-known properties of the CUE~\cite{mehta1991random}, the SFF admits what we call a moment expansion,
\begin{equation}
    K_{\mU}(L,t)\equiv 2^{L}\left[1+\sum_{n=1}^{L}\frac{\mathtt{r}_{n}(L,t)}{n!}\right],
\end{equation}
\begin{equation}
\mathtt{r}_{n}(L,t) = \int\prod_{j=1}^{n}\left[\cos(t\theta_{j})d\theta_{j}\right]\mathtt{R}_{n}(\theta_{1},\cdots,\theta_{n}),
\end{equation}
where $\mathtt{R}_{n}$ is the $n$-point single-particle correlation function of the CUE and $\mathtt{r}_{n}(L,t)$ is the $n$th moment.

Thus far, this procedure can be applied with simple modifications to any random matrix ensemble; for example, Ref.~\cite{Liao2020} implemented a similar analysis for the GUE. However, the CUE has special properties which allow for an exact derivation of~\eqref{eq:ExactSFF} for any choice of $L$. In particular, the computation of the moments can be mapped onto a combinatorial problem of hardcore particles in a one-dimensional system. Using this  representation, we have shown that the SFF satisfies the following factorization identity. Given a system size $L$ and a time $t$, write $L=Nt+r$ for some integers $N$ and $r\in\left[0,t-1\right]$. Then
\begin{equation}\label{eq:SFF_Factorization}
\begin{aligned}
    K_{\mU}(L,t) &= K_{\mU}(N+1,1)^{r}\times K_{\mU}(N,1)^{t-r}.
\end{aligned}
\end{equation}
Combined with a proof that $K_{\mU}(L,1) = L+1$, this implies the claimed result~\eqref{eq:ExactSFF}.

\lsec{Other circular ensembles} The procedure for generating random unitaries for the COE and CSE closely parallels the structure we employed for the CUE. One simply draws a unitary from the appropriate circular ensemble, computes its eigenvalues, and uses them to define $\mU$ in Eq.~\eqref{eq:Udef}. These ensembles exhibit correlated single-particle quasienergies that are linearly related to the many-body quasienergies, so our scaling arguments apply and we expect that
\begin{equation}\label{eq:scaling}
    \frac{\log K_{\mU}(L,t)}{L} = f(t/L)
\end{equation}
for an undetermined function $f$ which depends on the random matrix ensemble. To study this, we have developed transfer matrix methods which compute the SFF exactly. These methods are derived in detail in Ref.~\cite{OurPaper2}; here we present numerical results in Fig.~\ref{fig:K_COE_CSE}. Our numerics indicate that the ansatz~\eqref{eq:scaling} achieves a scaling collapse for both the COE and CSE models. We leave it to future work to identify the scaling functions.

\lsec{Comparison with non-interacting GUE fermions} While our main result~\eqref{eq:ExactSFF} holds only for the CUE, it serves as a useful baseline of comparison with other random matrix ensembles, particularly the GUE. Here we will provide a numerical comparison of their SFFs. To study the GUE, we follow the recipe of Ref.~\cite{Liao2020}. First, draw an $L\times L$ Hermitian matrix $h$ from the GUE and define the many-body fermion Hamiltonian
\begin{equation}
H = \sum_{ij}c_{i}^{\dagger}h_{ij}c_{j}\,.
\end{equation}
The SFF is
\begin{equation}
K_{\text{GUE}}(L,t) = \langle |\tr\; e^{-iHt}|^{2}\rangle_{h}\,,
\end{equation}
where $t\in\mathbb{R}$ and we have used notation analogous to our treatment of the CUE in Eq.~\eqref{eq:Kdef}.

Fig.~\ref{fig:Kt_GUE} shows a comparison between $K_{\text{GUE}}$ and the exact CUE result~\eqref{eq:ExactSFF} for a range of system sizes. In both cases we measure time in units of the relevant single-particle Heisenberg time, where $t_{H}^{1,\text{(CUE)}} = L$ and $t_{H}^{1,\text{(GUE)}} = (\pi/\sqrt{2})L$~\cite{Cipolloni2023}. At early times, the GUE results exhibit oscillations which relax towards a ramp that is quantitatively close to the exact CUE result (Fig.~\ref{fig:Kt_GUE} inset). This behavior persists up to times which are an $O(1)$ fraction of the single-particle Heisenberg time.

In the thermodynamic limit $L\to\infty$, the correlation and cluster functions of the GUE and CUE are known to agree asymptotically~\cite{mehta1991random}. Combined with the numerical results of Fig.~\ref{fig:Kt_GUE}, this motivates us to conjecture that the GUE form factors follows our exact result for the CUE in the thermodynamic limit up to an unknown $O(1)$ fraction of the many-body Heisenberg time. Provided that this conjecture holds, it significantly expands the utility of our results for the CUE as a benchmark for other models, such as chaotic many-body systems with weak interactions.


\lsec{Crossover from one-body to many-body chaos}
Away from the exactly solvable point defined by~\eqref{eq:Udef}, we expect that the SFF  grows linearly with time until reaching a plateau for $t\approx t_H\sim\mD$. To probe the crossover from exponential to linear growth of the SFF, we consider a kicked Ising model (KIM) whose one-step evolution is given by~\cite{Bertini2018}
\begin{equation}\label{eq:KIM_Floquet}
   \mU_\mathrm{KIM} = \exp\left[ -i  \sum_{i=1}^L \left(J Z_i Z_{i+1}+\frac{\theta_i}{2} Z_i \right)\right] \exp\left[-i g \sum_{i=1}^L X_i \right]
\end{equation}
where the $\theta_i$ are quasienergies of a random $L\times L$ CUE unitary, $X_i$ and $Z_i$ are Pauli matrices, and we have used periodic boundary conditions~\footnote{For the purpose of numerical applications, the reader should note that eigenvalue solvers often report results in a prescribed order, e.g., extremal eigenvalues first. In the presence of interactions, permutation invariance of sites should be enforced by hand and we have done this by randomizing the quasienergy order following each unitary diagonalization.}. We parameterize the couplings as
\begin{equation}
    J=g = \frac{\pi}{4}\tau.
\end{equation}
At the non-interacting point $\tau=0$, the KIM has the SFF~\eqref{eq:ExactSFF} and grows exponentially in time. For $\tau =1$, Ref.~\cite{Bertini2018} has studied $\mU_{\text{KIM}}$ without introducing correlations among the $\theta_i$ and argued that it is maximally chaotic. In the strongly interacting regime, correlations among $\theta_j$ are irrelevant and we expect $t$-linear growth of the SFF. More precisely, the SFF with $\tau=1$ is described by the many-body COE, for which the SFF is known,
\begin{align}\label{eq:K_MB_COE}
    &K_\mathrm{COE}(L,t)\notag\\
    &=
    \begin{cases}
        2 t - t \log(1+2t/t_H) & (t\le t_H) \\
        2t_H - t\log[(2t+t_H)/(2t-t_H)] & (t>t_H)\,.
    \end{cases}
\end{align}
In contrast to our result~\eqref{eq:ExactSFF}, this SFF grows linearly in $t$  and a scaling collapse is obtained by rescaling $t\to t/t_H$ and $K(L,t)\to K(L,t)/t_H$.

Fig.~\ref{fig:KIM_SFF} highlights the crossover from the one-body CUE to the many-body COE as $\tau$ increases from 0 to 1.
For small $\tau$, the SFF closely tracks the exact solution~\eqref{eq:ExactSFF}, suggesting that the one-body CUE behavior survives against small perturbations.
For larger interactions, $\tau\gtrsim 0.5$, the SFF first grows exponentially over a narrow time window before approaching and following the many-body COE curve.
For the maximally chaotic case $\tau=1$, the SFF closely follows the many-body COE following initial transients, which is consistent with Ref.~\cite{Bertini2018}. In the intermediate regime $\tau=0.6$, the SFF exhibits a short-time peak at $t\sim t_H^{(1)}=L$, which could be interpreted as a competition between the single-particle exponential ramp and the approach to the many-body COE universality.
These results suggest that the exponential growth of the SFF is robust to the introduction of interactions on sufficiently short time scales and serves as a persistent signal of single-particle chaos.

\lsec{Discussion}
In this Letter, we have constructed a set of many-body systems which are non-interacting but nonetheless exhibit level repulsion through the statistics of single-particle states. Crucially, the many-body SFF of these models can be computed exactly for any system size, offering insights into spectral statistics without using simplifying approximations or limits. While in our models the single-particle sector exhibits random matrix statistics by construction, we expect our results to also be relevant for chaotic non-interacting systems with short-range Hamiltonians, such as Anderson models in the delocalized regime.
Moreover, we have also demonstrated that the models under investigation provide a useful basis of comparison with interacting Hamiltonian systems. Interested readers are encouraged to consult our companion paper for proofs of technical claims~\cite{OurPaper2}.

A particularly interesting direction for future work lies in understanding the crossover between single-particle and many-body chaos. We have initiated such a study in the context of the kicked Ising model~\eqref{eq:KIM_Floquet} and it would be interesting to follow this up by searching for signatures of ergodicity breaking~\cite{sierant_lewenstein_23}.

As a final comment, we note that the introduction of weak interactions in Fig.~\ref{fig:KIM_SFF} produces interesting features such as late-time oscillations. It is an interesting problem to understand the physical meaning of these oscillations and to potentially re-interpret existing data in the literature with this question in mind~\cite{Dag2023}. While such signals could be interpreted as sampling noise in other contexts, our exact calculations can clearly differentiate between noise and real signal.

\begin{acknowledgments}
We acknowledge fruitful discussions with Thomas Scaffidi, Ceren Da{\u{g}}, Keisuke Fujii, and Kaoru Mizuta. M.O.F. acknowledges support from the Faculty of Science at the University of Victoria through Thomas E. Baker. L.V. acknowledges support from the Slovenian Research and Innovation Agency (ARIS), Research core funding Grants No.~P1-0044, N1-0273, J1-50005 and N1-0369, as well as the Consolidator Grant Boundary-101126364 of the
European Research Council (ERC).
T. N. I. was supported by JST PRESTO Grant No.~JPMJPR2112, JSPS KAKENHI Grant No.~JP21K13852, and the Boston University CMT visitors program.
This research was supported in part by the International Centre for Theoretical Sciences (ICTS) via participation in the program ``Stability of Quantum Matter in and out of Equilibrium at Various Scales'' (code: ICTS/SQMVS2024/01).
\end{acknowledgments}

\bibliography{Paper1Refs}

\begin{thebibliography}{59}%
\makeatletter
\providecommand \@ifxundefined [1]{%
 \@ifx{#1\undefined}
}%
\providecommand \@ifnum [1]{%
 \ifnum #1\expandafter \@firstoftwo
 \else \expandafter \@secondoftwo
 \fi
}%
\providecommand \@ifx [1]{%
 \ifx #1\expandafter \@firstoftwo
 \else \expandafter \@secondoftwo
 \fi
}%
\providecommand \natexlab [1]{#1}%
\providecommand \enquote  [1]{``#1''}%
\providecommand \bibnamefont  [1]{#1}%
\providecommand \bibfnamefont [1]{#1}%
\providecommand \citenamefont [1]{#1}%
\providecommand \href@noop [0]{\@secondoftwo}%
\providecommand \href [0]{\begingroup \@sanitize@url \@href}%
\providecommand \@href[1]{\@@startlink{#1}\@@href}%
\providecommand \@@href[1]{\endgroup#1\@@endlink}%
\providecommand \@sanitize@url [0]{\catcode `\\12\catcode `\$12\catcode `\&12\catcode `\#12\catcode `\^12\catcode `\_12\catcode `\%12\relax}%
\providecommand \@@startlink[1]{}%
\providecommand \@@endlink[0]{}%
\providecommand \url  [0]{\begingroup\@sanitize@url \@url }%
\providecommand \@url [1]{\endgroup\@href {#1}{\urlprefix }}%
\providecommand \urlprefix  [0]{URL }%
\providecommand \Eprint [0]{\href }%
\providecommand \doibase [0]{https://doi.org/}%
\providecommand \selectlanguage [0]{\@gobble}%
\providecommand \bibinfo  [0]{\@secondoftwo}%
\providecommand \bibfield  [0]{\@secondoftwo}%
\providecommand \translation [1]{[#1]}%
\providecommand \BibitemOpen [0]{}%
\providecommand \bibitemStop [0]{}%
\providecommand \bibitemNoStop [0]{.\EOS\space}%
\providecommand \EOS [0]{\spacefactor3000\relax}%
\providecommand \BibitemShut  [1]{\csname bibitem#1\endcsname}%
\let\auto@bib@innerbib\@empty
\bibitem [{\citenamefont {{Bloch}}\ \emph {et~al.}(2012)\citenamefont {{Bloch}}, \citenamefont {{Dalibard}},\ and\ \citenamefont {{Nascimb{\`e}ne}}}]{BlochQuantumSimulators}%
  \BibitemOpen
  \bibfield  {author} {\bibinfo {author} {\bibfnamefont {I.}~\bibnamefont {{Bloch}}}, \bibinfo {author} {\bibfnamefont {J.}~\bibnamefont {{Dalibard}}},\ and\ \bibinfo {author} {\bibfnamefont {S.}~\bibnamefont {{Nascimb{\`e}ne}}},\ }\bibfield  {title} {\bibinfo {title} {{Quantum simulations with ultracold quantum gases}},\ }\href {https://doi.org/10.1038/nphys2259} {\bibfield  {journal} {\bibinfo  {journal} {Nature Physics}\ }\textbf {\bibinfo {volume} {8}},\ \bibinfo {pages} {267} (\bibinfo {year} {2012})}\BibitemShut {NoStop}%
\bibitem [{\citenamefont {Schäfer}\ \emph {et~al.}(2020)\citenamefont {Schäfer}, \citenamefont {Fukuhara}, \citenamefont {Sugawa}, \citenamefont {Takasu},\ and\ \citenamefont {Takahashi}}]{QuantumSimulatorReview}%
  \BibitemOpen
  \bibfield  {author} {\bibinfo {author} {\bibfnamefont {F.}~\bibnamefont {Schäfer}}, \bibinfo {author} {\bibfnamefont {T.}~\bibnamefont {Fukuhara}}, \bibinfo {author} {\bibfnamefont {S.}~\bibnamefont {Sugawa}}, \bibinfo {author} {\bibfnamefont {Y.}~\bibnamefont {Takasu}},\ and\ \bibinfo {author} {\bibfnamefont {Y.}~\bibnamefont {Takahashi}},\ }\bibfield  {title} {\bibinfo {title} {Tools for quantum simulation with ultracold atoms in optical lattices},\ }\href {https://doi.org/10.1038/s42254-020-0195-3} {\bibfield  {journal} {\bibinfo  {journal} {Nature Reviews Physics}\ }\textbf {\bibinfo {volume} {2}},\ \bibinfo {pages} {411–425} (\bibinfo {year} {2020})}\BibitemShut {NoStop}%
\bibitem [{\citenamefont {{Georgescu}}\ \emph {et~al.}(2014)\citenamefont {{Georgescu}}, \citenamefont {{Ashhab}},\ and\ \citenamefont {{Nori}}}]{RMP_Simulators}%
  \BibitemOpen
  \bibfield  {author} {\bibinfo {author} {\bibfnamefont {I.~M.}\ \bibnamefont {{Georgescu}}}, \bibinfo {author} {\bibfnamefont {S.}~\bibnamefont {{Ashhab}}},\ and\ \bibinfo {author} {\bibfnamefont {F.}~\bibnamefont {{Nori}}},\ }\bibfield  {title} {\bibinfo {title} {{Quantum simulation}},\ }\href {https://doi.org/10.1103/RevModPhys.86.153} {\bibfield  {journal} {\bibinfo  {journal} {Reviews of Modern Physics}\ }\textbf {\bibinfo {volume} {86}},\ \bibinfo {pages} {153} (\bibinfo {year} {2014})}\BibitemShut {NoStop}%
\bibitem [{\citenamefont {Gross}\ and\ \citenamefont {Bloch}(2017)}]{Science_Simulators}%
  \BibitemOpen
  \bibfield  {author} {\bibinfo {author} {\bibfnamefont {C.}~\bibnamefont {Gross}}\ and\ \bibinfo {author} {\bibfnamefont {I.}~\bibnamefont {Bloch}},\ }\bibfield  {title} {\bibinfo {title} {Quantum simulations with ultracold atoms in optical lattices},\ }\href {https://doi.org/10.1126/science.aal3837} {\bibfield  {journal} {\bibinfo  {journal} {Science}\ }\textbf {\bibinfo {volume} {357}},\ \bibinfo {pages} {995} (\bibinfo {year} {2017})}\BibitemShut {NoStop}%
\bibitem [{\citenamefont {Mori}\ \emph {et~al.}(2018)\citenamefont {Mori}, \citenamefont {Ikeda}, \citenamefont {Kaminishi},\ and\ \citenamefont {Ueda}}]{ThermalizationOverview}%
  \BibitemOpen
  \bibfield  {author} {\bibinfo {author} {\bibfnamefont {T.}~\bibnamefont {Mori}}, \bibinfo {author} {\bibfnamefont {T.~N.}\ \bibnamefont {Ikeda}}, \bibinfo {author} {\bibfnamefont {E.}~\bibnamefont {Kaminishi}},\ and\ \bibinfo {author} {\bibfnamefont {M.}~\bibnamefont {Ueda}},\ }\bibfield  {title} {\bibinfo {title} {Thermalization and prethermalization in isolated quantum systems: a theoretical overview},\ }\href {https://doi.org/10.1088/1361-6455/aabcdf} {\bibfield  {journal} {\bibinfo  {journal} {Journal of Physics B: Atomic, Molecular and Optical Physics}\ }\textbf {\bibinfo {volume} {51}},\ \bibinfo {pages} {112001} (\bibinfo {year} {2018})}\BibitemShut {NoStop}%
\bibitem [{\citenamefont {Pandey}\ \emph {et~al.}(2020)\citenamefont {Pandey}, \citenamefont {Claeys}, \citenamefont {Campbell}, \citenamefont {Polkovnikov},\ and\ \citenamefont {Sels}}]{AdiabaticEigenstateDeformations}%
  \BibitemOpen
  \bibfield  {author} {\bibinfo {author} {\bibfnamefont {M.}~\bibnamefont {Pandey}}, \bibinfo {author} {\bibfnamefont {P.~W.}\ \bibnamefont {Claeys}}, \bibinfo {author} {\bibfnamefont {D.~K.}\ \bibnamefont {Campbell}}, \bibinfo {author} {\bibfnamefont {A.}~\bibnamefont {Polkovnikov}},\ and\ \bibinfo {author} {\bibfnamefont {D.}~\bibnamefont {Sels}},\ }\bibfield  {title} {\bibinfo {title} {Adiabatic eigenstate deformations as a sensitive probe for quantum chaos},\ }\href {https://doi.org/10.1103/PhysRevX.10.041017} {\bibfield  {journal} {\bibinfo  {journal} {Phys. Rev. X}\ }\textbf {\bibinfo {volume} {10}},\ \bibinfo {pages} {041017} (\bibinfo {year} {2020})}\BibitemShut {NoStop}%
\bibitem [{\citenamefont {Lim}\ \emph {et~al.}(2024)\citenamefont {Lim}, \citenamefont {Matirko}, \citenamefont {Polkovnikov},\ and\ \citenamefont {Flynn}}]{DefiningClassicalQuantumChaos}%
  \BibitemOpen
  \bibfield  {author} {\bibinfo {author} {\bibfnamefont {C.}~\bibnamefont {Lim}}, \bibinfo {author} {\bibfnamefont {K.}~\bibnamefont {Matirko}}, \bibinfo {author} {\bibfnamefont {A.}~\bibnamefont {Polkovnikov}},\ and\ \bibinfo {author} {\bibfnamefont {M.~O.}\ \bibnamefont {Flynn}},\ }\href {https://arxiv.org/abs/2401.01927} {\bibinfo {title} {Defining classical and quantum chaos through adiabatic transformations}} (\bibinfo {year} {2024}),\ \Eprint {https://arxiv.org/abs/2401.01927} {arXiv:2401.01927 [cond-mat.stat-mech]} \BibitemShut {NoStop}%
\bibitem [{\citenamefont {Parker}\ \emph {et~al.}(2019)\citenamefont {Parker}, \citenamefont {Cao}, \citenamefont {Avdoshkin}, \citenamefont {Scaffidi},\ and\ \citenamefont {Altman}}]{OperatorGrowthHypothesis}%
  \BibitemOpen
  \bibfield  {author} {\bibinfo {author} {\bibfnamefont {D.~E.}\ \bibnamefont {Parker}}, \bibinfo {author} {\bibfnamefont {X.}~\bibnamefont {Cao}}, \bibinfo {author} {\bibfnamefont {A.}~\bibnamefont {Avdoshkin}}, \bibinfo {author} {\bibfnamefont {T.}~\bibnamefont {Scaffidi}},\ and\ \bibinfo {author} {\bibfnamefont {E.}~\bibnamefont {Altman}},\ }\bibfield  {title} {\bibinfo {title} {A universal operator growth hypothesis},\ }\href {https://doi.org/10.1103/PhysRevX.9.041017} {\bibfield  {journal} {\bibinfo  {journal} {Phys. Rev. X}\ }\textbf {\bibinfo {volume} {9}},\ \bibinfo {pages} {041017} (\bibinfo {year} {2019})}\BibitemShut {NoStop}%
\bibitem [{\citenamefont {Deutsch}(1991)}]{DeutschStatMech}%
  \BibitemOpen
  \bibfield  {author} {\bibinfo {author} {\bibfnamefont {J.~M.}\ \bibnamefont {Deutsch}},\ }\bibfield  {title} {\bibinfo {title} {Quantum statistical mechanics in a closed system},\ }\href {https://doi.org/10.1103/PhysRevA.43.2046} {\bibfield  {journal} {\bibinfo  {journal} {Phys. Rev. A}\ }\textbf {\bibinfo {volume} {43}},\ \bibinfo {pages} {2046} (\bibinfo {year} {1991})}\BibitemShut {NoStop}%
\bibitem [{\citenamefont {Srednicki}(1994)}]{Srednicki1}%
  \BibitemOpen
  \bibfield  {author} {\bibinfo {author} {\bibfnamefont {M.}~\bibnamefont {Srednicki}},\ }\bibfield  {title} {\bibinfo {title} {Chaos and quantum thermalization},\ }\href {https://doi.org/10.1103/PhysRevE.50.888} {\bibfield  {journal} {\bibinfo  {journal} {Phys. Rev. E}\ }\textbf {\bibinfo {volume} {50}},\ \bibinfo {pages} {888} (\bibinfo {year} {1994})}\BibitemShut {NoStop}%
\bibitem [{\citenamefont {Srednicki}(1999)}]{Srednicki2}%
  \BibitemOpen
  \bibfield  {author} {\bibinfo {author} {\bibfnamefont {M.}~\bibnamefont {Srednicki}},\ }\bibfield  {title} {\bibinfo {title} {The approach to thermal equilibrium in quantized chaotic systems},\ }\href {https://doi.org/10.1088/0305-4470/32/7/007} {\bibfield  {journal} {\bibinfo  {journal} {Journal of Physics A: Mathematical and General}\ }\textbf {\bibinfo {volume} {32}},\ \bibinfo {pages} {1163–1175} (\bibinfo {year} {1999})}\BibitemShut {NoStop}%
\bibitem [{\citenamefont {Srednicki}(1996)}]{Srenicki3}%
  \BibitemOpen
  \bibfield  {author} {\bibinfo {author} {\bibfnamefont {M.}~\bibnamefont {Srednicki}},\ }\bibfield  {title} {\bibinfo {title} {Thermal fluctuations in quantized chaotic systems},\ }\href {https://doi.org/10.1088/0305-4470/29/4/003} {\bibfield  {journal} {\bibinfo  {journal} {Journal of Physics A: Mathematical and General}\ }\textbf {\bibinfo {volume} {29}},\ \bibinfo {pages} {L75–L79} (\bibinfo {year} {1996})}\BibitemShut {NoStop}%
\bibitem [{\citenamefont {Rigol}\ \emph {et~al.}(2008)\citenamefont {Rigol}, \citenamefont {Dunjko},\ and\ \citenamefont {Olshanii}}]{rigol_dunjko_08}%
  \BibitemOpen
  \bibfield  {author} {\bibinfo {author} {\bibfnamefont {M.}~\bibnamefont {Rigol}}, \bibinfo {author} {\bibfnamefont {V.}~\bibnamefont {Dunjko}},\ and\ \bibinfo {author} {\bibfnamefont {M.}~\bibnamefont {Olshanii}},\ }\bibfield  {title} {\bibinfo {title} {Thermalization and its mechanism for generic isolated quantum systems},\ }\href {https://doi.org/10.1038/nature06838} {\bibfield  {journal} {\bibinfo  {journal} {Nature (London)}\ }\textbf {\bibinfo {volume} {452}},\ \bibinfo {pages} {854} (\bibinfo {year} {2008})}\BibitemShut {NoStop}%
\bibitem [{\citenamefont {D’Alessio}\ \emph {et~al.}(2016)\citenamefont {D’Alessio}, \citenamefont {Kafri}, \citenamefont {Polkovnikov},\ and\ \citenamefont {Rigol}}]{ETHReview}%
  \BibitemOpen
  \bibfield  {author} {\bibinfo {author} {\bibfnamefont {L.}~\bibnamefont {D’Alessio}}, \bibinfo {author} {\bibfnamefont {Y.}~\bibnamefont {Kafri}}, \bibinfo {author} {\bibfnamefont {A.}~\bibnamefont {Polkovnikov}},\ and\ \bibinfo {author} {\bibfnamefont {M.}~\bibnamefont {Rigol}},\ }\bibfield  {title} {\bibinfo {title} {From quantum chaos and eigenstate thermalization to statistical mechanics and thermodynamics},\ }\href {https://doi.org/10.1080/00018732.2016.1198134} {\bibfield  {journal} {\bibinfo  {journal} {Advances in Physics}\ }\textbf {\bibinfo {volume} {65}},\ \bibinfo {pages} {239–362} (\bibinfo {year} {2016})}\BibitemShut {NoStop}%
\bibitem [{\citenamefont {Wigner}(1955)}]{Wigner1}%
  \BibitemOpen
  \bibfield  {author} {\bibinfo {author} {\bibfnamefont {E.~P.}\ \bibnamefont {Wigner}},\ }\bibfield  {title} {\bibinfo {title} {Characteristic vectors of bordered matrices with infinite dimensions},\ }\href {http://www.jstor.org/stable/1970079} {\bibfield  {journal} {\bibinfo  {journal} {Annals of Mathematics}\ }\textbf {\bibinfo {volume} {62}},\ \bibinfo {pages} {548} (\bibinfo {year} {1955})}\BibitemShut {NoStop}%
\bibitem [{\citenamefont {Wigner}(1957)}]{Wigner2}%
  \BibitemOpen
  \bibfield  {author} {\bibinfo {author} {\bibfnamefont {E.~P.}\ \bibnamefont {Wigner}},\ }\bibfield  {title} {\bibinfo {title} {Characteristics vectors of bordered matrices with infinite dimensions ii},\ }\href {http://www.jstor.org/stable/1969956} {\bibfield  {journal} {\bibinfo  {journal} {Annals of Mathematics}\ }\textbf {\bibinfo {volume} {65}},\ \bibinfo {pages} {203} (\bibinfo {year} {1957})}\BibitemShut {NoStop}%
\bibitem [{\citenamefont {Wigner}(1958)}]{Wigner3}%
  \BibitemOpen
  \bibfield  {author} {\bibinfo {author} {\bibfnamefont {E.~P.}\ \bibnamefont {Wigner}},\ }\bibfield  {title} {\bibinfo {title} {On the distribution of the roots of certain symmetric matrices},\ }\href {http://www.jstor.org/stable/1970008} {\bibfield  {journal} {\bibinfo  {journal} {Annals of Mathematics}\ }\textbf {\bibinfo {volume} {67}},\ \bibinfo {pages} {325} (\bibinfo {year} {1958})}\BibitemShut {NoStop}%
\bibitem [{\citenamefont {Alhassid}(2000)}]{RMTreviewOne}%
  \BibitemOpen
  \bibfield  {author} {\bibinfo {author} {\bibfnamefont {Y.}~\bibnamefont {Alhassid}},\ }\bibfield  {title} {\bibinfo {title} {The statistical theory of quantum dots},\ }\href {https://doi.org/10.1103/RevModPhys.72.895} {\bibfield  {journal} {\bibinfo  {journal} {Rev. Mod. Phys.}\ }\textbf {\bibinfo {volume} {72}},\ \bibinfo {pages} {895} (\bibinfo {year} {2000})}\BibitemShut {NoStop}%
\bibitem [{\citenamefont {Kravtsov}(2012)}]{RMTReview2}%
  \BibitemOpen
  \bibfield  {author} {\bibinfo {author} {\bibfnamefont {V.~E.}\ \bibnamefont {Kravtsov}},\ }\href@noop {} {\bibinfo {title} {Random matrix theory: Wigner-dyson statistics and beyond. (lecture notes of a course given at sissa (trieste, italy))}} (\bibinfo {year} {2012}),\ \Eprint {https://arxiv.org/abs/0911.0639} {arXiv:0911.0639 [cond-mat.dis-nn]} \BibitemShut {NoStop}%
\bibitem [{\citenamefont {Sierant}\ \emph {et~al.}()\citenamefont {Sierant}, \citenamefont {Lewenstein}, \citenamefont {Scardicchio}, \citenamefont {Vidmar},\ and\ \citenamefont {Zakrzewski}}]{sierant_lewenstein_24}%
  \BibitemOpen
  \bibfield  {author} {\bibinfo {author} {\bibfnamefont {P.}~\bibnamefont {Sierant}}, \bibinfo {author} {\bibfnamefont {M.}~\bibnamefont {Lewenstein}}, \bibinfo {author} {\bibfnamefont {A.}~\bibnamefont {Scardicchio}}, \bibinfo {author} {\bibfnamefont {L.}~\bibnamefont {Vidmar}},\ and\ \bibinfo {author} {\bibfnamefont {J.}~\bibnamefont {Zakrzewski}},\ }\href {https://arxiv.org/abs/2403.07111} {\bibinfo {title} {Many-body localization in the age of classical computing}},\ \bibinfo {howpublished} {arXiv:2403.07111}\BibitemShut {NoStop}%
\bibitem [{\citenamefont {Collins}\ and\ \citenamefont {Nechita}(2015)}]{RMTBlock1}%
  \BibitemOpen
  \bibfield  {author} {\bibinfo {author} {\bibfnamefont {B.}~\bibnamefont {Collins}}\ and\ \bibinfo {author} {\bibfnamefont {I.}~\bibnamefont {Nechita}},\ }\bibfield  {title} {\bibinfo {title} {Random matrix techniques in quantum information theory},\ }\bibfield  {journal} {\bibinfo  {journal} {Journal of Mathematical Physics}\ }\textbf {\bibinfo {volume} {57}},\ \href {https://doi.org/10.1063/1.4936880} {10.1063/1.4936880} (\bibinfo {year} {2015})\BibitemShut {NoStop}%
\bibitem [{\citenamefont {Guhr}\ \emph {et~al.}(1998)\citenamefont {Guhr}, \citenamefont {Müller–Groeling},\ and\ \citenamefont {Weidenmüller}}]{RMTBlock2}%
  \BibitemOpen
  \bibfield  {author} {\bibinfo {author} {\bibfnamefont {T.}~\bibnamefont {Guhr}}, \bibinfo {author} {\bibfnamefont {A.}~\bibnamefont {Müller–Groeling}},\ and\ \bibinfo {author} {\bibfnamefont {H.~A.}\ \bibnamefont {Weidenmüller}},\ }\bibfield  {title} {\bibinfo {title} {Random-matrix theories in quantum physics: common concepts},\ }\href {https://doi.org/10.1016/s0370-1573(97)00088-4} {\bibfield  {journal} {\bibinfo  {journal} {Physics Reports}\ }\textbf {\bibinfo {volume} {299}},\ \bibinfo {pages} {189–425} (\bibinfo {year} {1998})}\BibitemShut {NoStop}%
\bibitem [{\citenamefont {Roberts}\ and\ \citenamefont {Yoshida}(2017)}]{RMTBlock3}%
  \BibitemOpen
  \bibfield  {author} {\bibinfo {author} {\bibfnamefont {D.~A.}\ \bibnamefont {Roberts}}\ and\ \bibinfo {author} {\bibfnamefont {B.}~\bibnamefont {Yoshida}},\ }\bibfield  {title} {\bibinfo {title} {Chaos and complexity by design},\ }\bibfield  {journal} {\bibinfo  {journal} {Journal of High Energy Physics}\ }\textbf {\bibinfo {volume} {2017}},\ \href {https://doi.org/10.1007/jhep04(2017)121} {10.1007/jhep04(2017)121} (\bibinfo {year} {2017})\BibitemShut {NoStop}%
\bibitem [{\citenamefont {McDonald}\ and\ \citenamefont {Kaufman}(1979)}]{Billiards1}%
  \BibitemOpen
  \bibfield  {author} {\bibinfo {author} {\bibfnamefont {S.~W.}\ \bibnamefont {McDonald}}\ and\ \bibinfo {author} {\bibfnamefont {A.~N.}\ \bibnamefont {Kaufman}},\ }\bibfield  {title} {\bibinfo {title} {Spectrum and eigenfunctions for a hamiltonian with stochastic trajectories},\ }\href {https://doi.org/10.1103/PhysRevLett.42.1189} {\bibfield  {journal} {\bibinfo  {journal} {Phys. Rev. Lett.}\ }\textbf {\bibinfo {volume} {42}},\ \bibinfo {pages} {1189} (\bibinfo {year} {1979})}\BibitemShut {NoStop}%
\bibitem [{\citenamefont {Rozenbaum}\ \emph {et~al.}(2019)\citenamefont {Rozenbaum}, \citenamefont {Ganeshan},\ and\ \citenamefont {Galitski}}]{Billiards2}%
  \BibitemOpen
  \bibfield  {author} {\bibinfo {author} {\bibfnamefont {E.~B.}\ \bibnamefont {Rozenbaum}}, \bibinfo {author} {\bibfnamefont {S.}~\bibnamefont {Ganeshan}},\ and\ \bibinfo {author} {\bibfnamefont {V.}~\bibnamefont {Galitski}},\ }\bibfield  {title} {\bibinfo {title} {Universal level statistics of the out-of-time-ordered operator},\ }\href {https://doi.org/10.1103/PhysRevB.100.035112} {\bibfield  {journal} {\bibinfo  {journal} {Phys. Rev. B}\ }\textbf {\bibinfo {volume} {100}},\ \bibinfo {pages} {035112} (\bibinfo {year} {2019})}\BibitemShut {NoStop}%
\bibitem [{\citenamefont {Berry}\ and\ \citenamefont {Tabor}(1977)}]{Billiard3}%
  \BibitemOpen
  \bibfield  {author} {\bibinfo {author} {\bibfnamefont {M.~V.}\ \bibnamefont {Berry}}\ and\ \bibinfo {author} {\bibfnamefont {M.}~\bibnamefont {Tabor}},\ }\bibfield  {title} {\bibinfo {title} {Level clustering in the regular spectrum},\ }\href {https://doi.org/10.1098/rspa.1977.0140} {\bibfield  {journal} {\bibinfo  {journal} {Proceedings of the Royal Society of London A: Mathematical, Physical and Engineering Sciences}\ }\textbf {\bibinfo {volume} {356}},\ \bibinfo {pages} {375} (\bibinfo {year} {1977})}\BibitemShut {NoStop}%
\bibitem [{\citenamefont {Bohigas}\ \emph {et~al.}(1984)\citenamefont {Bohigas}, \citenamefont {Giannoni},\ and\ \citenamefont {Schmit}}]{Billiard4}%
  \BibitemOpen
  \bibfield  {author} {\bibinfo {author} {\bibfnamefont {O.}~\bibnamefont {Bohigas}}, \bibinfo {author} {\bibfnamefont {M.~J.}\ \bibnamefont {Giannoni}},\ and\ \bibinfo {author} {\bibfnamefont {C.}~\bibnamefont {Schmit}},\ }\bibfield  {title} {\bibinfo {title} {Characterization of chaotic quantum spectra and universality of level fluctuation laws},\ }\href {https://doi.org/10.1103/PhysRevLett.52.1} {\bibfield  {journal} {\bibinfo  {journal} {Phys. Rev. Lett.}\ }\textbf {\bibinfo {volume} {52}},\ \bibinfo {pages} {1} (\bibinfo {year} {1984})}\BibitemShut {NoStop}%
\bibitem [{\citenamefont {Kos}\ \emph {et~al.}(2018)\citenamefont {Kos}, \citenamefont {Ljubotina},\ and\ \citenamefont {Prosen}}]{Kos18}%
  \BibitemOpen
  \bibfield  {author} {\bibinfo {author} {\bibfnamefont {P.}~\bibnamefont {Kos}}, \bibinfo {author} {\bibfnamefont {M.}~\bibnamefont {Ljubotina}},\ and\ \bibinfo {author} {\bibfnamefont {T.}~\bibnamefont {Prosen}},\ }\bibfield  {title} {\bibinfo {title} {Many-body quantum chaos: Analytic connection to random matrix theory},\ }\href {https://doi.org/10.1103/PhysRevX.8.021062} {\bibfield  {journal} {\bibinfo  {journal} {Phys. Rev. X}\ }\textbf {\bibinfo {volume} {8}},\ \bibinfo {pages} {021062} (\bibinfo {year} {2018})}\BibitemShut {NoStop}%
\bibitem [{\citenamefont {Bertini}\ \emph {et~al.}(2018)\citenamefont {Bertini}, \citenamefont {Kos},\ and\ \citenamefont {Prosen}}]{Bertini2018}%
  \BibitemOpen
  \bibfield  {author} {\bibinfo {author} {\bibfnamefont {B.}~\bibnamefont {Bertini}}, \bibinfo {author} {\bibfnamefont {P.}~\bibnamefont {Kos}},\ and\ \bibinfo {author} {\bibfnamefont {T.}~\bibnamefont {Prosen}},\ }\bibfield  {title} {\bibinfo {title} {Exact spectral form factor in a minimal model of many-body quantum chaos},\ }\href {https://doi.org/10.1103/PhysRevLett.121.264101} {\bibfield  {journal} {\bibinfo  {journal} {Phys. Rev. Lett.}\ }\textbf {\bibinfo {volume} {121}},\ \bibinfo {pages} {264101} (\bibinfo {year} {2018})}\BibitemShut {NoStop}%
\bibitem [{\citenamefont {Chan}\ \emph {et~al.}(2018{\natexlab{a}})\citenamefont {Chan}, \citenamefont {De~Luca},\ and\ \citenamefont {Chalker}}]{ChanMinimalModel}%
  \BibitemOpen
  \bibfield  {author} {\bibinfo {author} {\bibfnamefont {A.}~\bibnamefont {Chan}}, \bibinfo {author} {\bibfnamefont {A.}~\bibnamefont {De~Luca}},\ and\ \bibinfo {author} {\bibfnamefont {J.~T.}\ \bibnamefont {Chalker}},\ }\bibfield  {title} {\bibinfo {title} {Solution of a minimal model for many-body quantum chaos},\ }\href {https://doi.org/10.1103/PhysRevX.8.041019} {\bibfield  {journal} {\bibinfo  {journal} {Phys. Rev. X}\ }\textbf {\bibinfo {volume} {8}},\ \bibinfo {pages} {041019} (\bibinfo {year} {2018}{\natexlab{a}})}\BibitemShut {NoStop}%
\bibitem [{\citenamefont {Chan}\ \emph {et~al.}(2018{\natexlab{b}})\citenamefont {Chan}, \citenamefont {De~Luca},\ and\ \citenamefont {Chalker}}]{Chan2018}%
  \BibitemOpen
  \bibfield  {author} {\bibinfo {author} {\bibfnamefont {A.}~\bibnamefont {Chan}}, \bibinfo {author} {\bibfnamefont {A.}~\bibnamefont {De~Luca}},\ and\ \bibinfo {author} {\bibfnamefont {J.~T.}\ \bibnamefont {Chalker}},\ }\bibfield  {title} {\bibinfo {title} {Spectral statistics in spatially extended chaotic quantum many-body systems},\ }\href {https://doi.org/10.1103/PhysRevLett.121.060601} {\bibfield  {journal} {\bibinfo  {journal} {Phys. Rev. Lett.}\ }\textbf {\bibinfo {volume} {121}},\ \bibinfo {pages} {060601} (\bibinfo {year} {2018}{\natexlab{b}})}\BibitemShut {NoStop}%
\bibitem [{\citenamefont {Liu}(2018)}]{Liu2018}%
  \BibitemOpen
  \bibfield  {author} {\bibinfo {author} {\bibfnamefont {J.}~\bibnamefont {Liu}},\ }\bibfield  {title} {\bibinfo {title} {Spectral form factors and late time quantum chaos},\ }\href {https://doi.org/10.1103/PhysRevD.98.086026} {\bibfield  {journal} {\bibinfo  {journal} {Phys. Rev. D}\ }\textbf {\bibinfo {volume} {98}},\ \bibinfo {pages} {086026} (\bibinfo {year} {2018})}\BibitemShut {NoStop}%
\bibitem [{\citenamefont {Friedman}\ \emph {et~al.}(2019)\citenamefont {Friedman}, \citenamefont {Chan}, \citenamefont {De~Luca},\ and\ \citenamefont {Chalker}}]{ChanConservedCharge}%
  \BibitemOpen
  \bibfield  {author} {\bibinfo {author} {\bibfnamefont {A.~J.}\ \bibnamefont {Friedman}}, \bibinfo {author} {\bibfnamefont {A.}~\bibnamefont {Chan}}, \bibinfo {author} {\bibfnamefont {A.}~\bibnamefont {De~Luca}},\ and\ \bibinfo {author} {\bibfnamefont {J.~T.}\ \bibnamefont {Chalker}},\ }\bibfield  {title} {\bibinfo {title} {Spectral statistics and many-body quantum chaos with conserved charge},\ }\href {https://doi.org/10.1103/PhysRevLett.123.210603} {\bibfield  {journal} {\bibinfo  {journal} {Phys. Rev. Lett.}\ }\textbf {\bibinfo {volume} {123}},\ \bibinfo {pages} {210603} (\bibinfo {year} {2019})}\BibitemShut {NoStop}%
\bibitem [{\citenamefont {\v{S}untajs}\ \emph {et~al.}(2020)\citenamefont {\v{S}untajs}, \citenamefont {Bon\v{c}a}, \citenamefont {Prosen},\ and\ \citenamefont {Vidmar}}]{suntajs_bonca_20a}%
  \BibitemOpen
  \bibfield  {author} {\bibinfo {author} {\bibfnamefont {J.}~\bibnamefont {\v{S}untajs}}, \bibinfo {author} {\bibfnamefont {J.}~\bibnamefont {Bon\v{c}a}}, \bibinfo {author} {\bibfnamefont {T.}~\bibnamefont {Prosen}},\ and\ \bibinfo {author} {\bibfnamefont {L.}~\bibnamefont {Vidmar}},\ }\bibfield  {title} {\bibinfo {title} {Quantum chaos challenges many-body localization},\ }\href {https://doi.org/10.1103/PhysRevE.102.062144} {\bibfield  {journal} {\bibinfo  {journal} {Phys. Rev. E}\ }\textbf {\bibinfo {volume} {102}},\ \bibinfo {pages} {062144} (\bibinfo {year} {2020})}\BibitemShut {NoStop}%
\bibitem [{\citenamefont {Sierant}\ \emph {et~al.}(2020)\citenamefont {Sierant}, \citenamefont {Delande},\ and\ \citenamefont {Zakrzewski}}]{sierant_delande_20}%
  \BibitemOpen
  \bibfield  {author} {\bibinfo {author} {\bibfnamefont {P.}~\bibnamefont {Sierant}}, \bibinfo {author} {\bibfnamefont {D.}~\bibnamefont {Delande}},\ and\ \bibinfo {author} {\bibfnamefont {J.}~\bibnamefont {Zakrzewski}},\ }\bibfield  {title} {\bibinfo {title} {{Thouless Time Analysis of Anderson and Many-Body Localization Transitions}},\ }\href {https://doi.org/10.1103/PhysRevLett.124.186601} {\bibfield  {journal} {\bibinfo  {journal} {Phys. Rev. Lett.}\ }\textbf {\bibinfo {volume} {124}},\ \bibinfo {pages} {186601} (\bibinfo {year} {2020})}\BibitemShut {NoStop}%
\bibitem [{\citenamefont {\v{S}untajs}\ \emph {et~al.}(2021)\citenamefont {\v{S}untajs}, \citenamefont {Prosen},\ and\ \citenamefont {Vidmar}}]{suntajs_prosen_21}%
  \BibitemOpen
  \bibfield  {author} {\bibinfo {author} {\bibfnamefont {J.}~\bibnamefont {\v{S}untajs}}, \bibinfo {author} {\bibfnamefont {T.}~\bibnamefont {Prosen}},\ and\ \bibinfo {author} {\bibfnamefont {L.}~\bibnamefont {Vidmar}},\ }\bibfield  {title} {\bibinfo {title} {{Spectral properties of three-dimensional Anderson model}},\ }\href {https://doi.org/https://doi.org/10.1016/j.aop.2021.168469} {\bibfield  {journal} {\bibinfo  {journal} {Ann. Phys. (Amsterdam)}\ }\textbf {\bibinfo {volume} {435}},\ \bibinfo {pages} {168469} (\bibinfo {year} {2021})}\BibitemShut {NoStop}%
\bibitem [{\citenamefont {Vasilyev}\ \emph {et~al.}(2020)\citenamefont {Vasilyev}, \citenamefont {Grankin}, \citenamefont {Baranov}, \citenamefont {Sieberer},\ and\ \citenamefont {Zoller}}]{Vasilyev20}%
  \BibitemOpen
  \bibfield  {author} {\bibinfo {author} {\bibfnamefont {D.~V.}\ \bibnamefont {Vasilyev}}, \bibinfo {author} {\bibfnamefont {A.}~\bibnamefont {Grankin}}, \bibinfo {author} {\bibfnamefont {M.~A.}\ \bibnamefont {Baranov}}, \bibinfo {author} {\bibfnamefont {L.~M.}\ \bibnamefont {Sieberer}},\ and\ \bibinfo {author} {\bibfnamefont {P.}~\bibnamefont {Zoller}},\ }\bibfield  {title} {\bibinfo {title} {Monitoring quantum simulators via quantum nondemolition couplings to atomic clock qubits},\ }\href {https://doi.org/10.1103/PRXQuantum.1.020302} {\bibfield  {journal} {\bibinfo  {journal} {PRX Quantum}\ }\textbf {\bibinfo {volume} {1}},\ \bibinfo {pages} {020302} (\bibinfo {year} {2020})}\BibitemShut {NoStop}%
\bibitem [{\citenamefont {Prakash}\ \emph {et~al.}(2021)\citenamefont {Prakash}, \citenamefont {Pixley},\ and\ \citenamefont {Kulkarni}}]{Prakash21}%
  \BibitemOpen
  \bibfield  {author} {\bibinfo {author} {\bibfnamefont {A.}~\bibnamefont {Prakash}}, \bibinfo {author} {\bibfnamefont {J.~H.}\ \bibnamefont {Pixley}},\ and\ \bibinfo {author} {\bibfnamefont {M.}~\bibnamefont {Kulkarni}},\ }\bibfield  {title} {\bibinfo {title} {Universal spectral form factor for many-body localization},\ }\href {https://doi.org/10.1103/PhysRevResearch.3.L012019} {\bibfield  {journal} {\bibinfo  {journal} {Phys. Rev. Res.}\ }\textbf {\bibinfo {volume} {3}},\ \bibinfo {pages} {L012019} (\bibinfo {year} {2021})}\BibitemShut {NoStop}%
\bibitem [{\citenamefont {Roy}\ and\ \citenamefont {Prosen}(2020)}]{Dibyendu1}%
  \BibitemOpen
  \bibfield  {author} {\bibinfo {author} {\bibfnamefont {D.}~\bibnamefont {Roy}}\ and\ \bibinfo {author} {\bibfnamefont {T.~c.~v.}\ \bibnamefont {Prosen}},\ }\bibfield  {title} {\bibinfo {title} {Random matrix spectral form factor in kicked interacting fermionic chains},\ }\href {https://doi.org/10.1103/PhysRevE.102.060202} {\bibfield  {journal} {\bibinfo  {journal} {Phys. Rev. E}\ }\textbf {\bibinfo {volume} {102}},\ \bibinfo {pages} {060202} (\bibinfo {year} {2020})}\BibitemShut {NoStop}%
\bibitem [{\citenamefont {Roy}\ \emph {et~al.}(2022)\citenamefont {Roy}, \citenamefont {Mishra},\ and\ \citenamefont {Prosen}}]{Dibyendu2}%
  \BibitemOpen
  \bibfield  {author} {\bibinfo {author} {\bibfnamefont {D.}~\bibnamefont {Roy}}, \bibinfo {author} {\bibfnamefont {D.}~\bibnamefont {Mishra}},\ and\ \bibinfo {author} {\bibfnamefont {T.~c.~v.}\ \bibnamefont {Prosen}},\ }\bibfield  {title} {\bibinfo {title} {Spectral form factor in a minimal bosonic model of many-body quantum chaos},\ }\href {https://doi.org/10.1103/PhysRevE.106.024208} {\bibfield  {journal} {\bibinfo  {journal} {Phys. Rev. E}\ }\textbf {\bibinfo {volume} {106}},\ \bibinfo {pages} {024208} (\bibinfo {year} {2022})}\BibitemShut {NoStop}%
\bibitem [{\citenamefont {Liao}\ and\ \citenamefont {Galitski}(2022)}]{Liao2}%
  \BibitemOpen
  \bibfield  {author} {\bibinfo {author} {\bibfnamefont {Y.}~\bibnamefont {Liao}}\ and\ \bibinfo {author} {\bibfnamefont {V.}~\bibnamefont {Galitski}},\ }\bibfield  {title} {\bibinfo {title} {Universal dephasing mechanism of many-body quantum chaos},\ }\href {https://doi.org/10.1103/PhysRevResearch.4.L012037} {\bibfield  {journal} {\bibinfo  {journal} {Phys. Rev. Res.}\ }\textbf {\bibinfo {volume} {4}},\ \bibinfo {pages} {L012037} (\bibinfo {year} {2022})}\BibitemShut {NoStop}%
\bibitem [{\citenamefont {Šuntajs}\ and\ \citenamefont {Vidmar}(2022)}]{ErgodicityBreakingZeroDimensions}%
  \BibitemOpen
  \bibfield  {author} {\bibinfo {author} {\bibfnamefont {J.}~\bibnamefont {Šuntajs}}\ and\ \bibinfo {author} {\bibfnamefont {L.}~\bibnamefont {Vidmar}},\ }\bibfield  {title} {\bibinfo {title} {Ergodicity breaking transition in zero dimensions},\ }\href {http://dx.doi.org/10.1103/PhysRevLett.129.060602} {\bibfield  {journal} {\bibinfo  {journal} {Phys. Rev. Lett.}\ }\textbf {\bibinfo {volume} {129}} (\bibinfo {year} {2022})}\BibitemShut {NoStop}%
\bibitem [{\citenamefont {Joshi}\ \emph {et~al.}(2022)\citenamefont {Joshi}, \citenamefont {Elben}, \citenamefont {Vikram}, \citenamefont {Vermersch}, \citenamefont {Galitski},\ and\ \citenamefont {Zoller}}]{Joshi22}%
  \BibitemOpen
  \bibfield  {author} {\bibinfo {author} {\bibfnamefont {L.~K.}\ \bibnamefont {Joshi}}, \bibinfo {author} {\bibfnamefont {A.}~\bibnamefont {Elben}}, \bibinfo {author} {\bibfnamefont {A.}~\bibnamefont {Vikram}}, \bibinfo {author} {\bibfnamefont {B.}~\bibnamefont {Vermersch}}, \bibinfo {author} {\bibfnamefont {V.}~\bibnamefont {Galitski}},\ and\ \bibinfo {author} {\bibfnamefont {P.}~\bibnamefont {Zoller}},\ }\bibfield  {title} {\bibinfo {title} {Probing many-body quantum chaos with quantum simulators},\ }\href {https://doi.org/10.1103/PhysRevX.12.011018} {\bibfield  {journal} {\bibinfo  {journal} {Phys. Rev. X}\ }\textbf {\bibinfo {volume} {12}},\ \bibinfo {pages} {011018} (\bibinfo {year} {2022})}\BibitemShut {NoStop}%
\bibitem [{\citenamefont {Winer}\ and\ \citenamefont {Swingle}(2022)}]{WinerHydroSFF}%
  \BibitemOpen
  \bibfield  {author} {\bibinfo {author} {\bibfnamefont {M.}~\bibnamefont {Winer}}\ and\ \bibinfo {author} {\bibfnamefont {B.}~\bibnamefont {Swingle}},\ }\bibfield  {title} {\bibinfo {title} {Hydrodynamic theory of the connected spectral form factor},\ }\href {https://doi.org/10.1103/PhysRevX.12.021009} {\bibfield  {journal} {\bibinfo  {journal} {Phys. Rev. X}\ }\textbf {\bibinfo {volume} {12}},\ \bibinfo {pages} {021009} (\bibinfo {year} {2022})}\BibitemShut {NoStop}%
\bibitem [{\citenamefont {Liu}\ \emph {et~al.}(2017)\citenamefont {Liu}, \citenamefont {Nowak},\ and\ \citenamefont {Zahed}}]{Liu17}%
  \BibitemOpen
  \bibfield  {author} {\bibinfo {author} {\bibfnamefont {Y.}~\bibnamefont {Liu}}, \bibinfo {author} {\bibfnamefont {M.~A.}\ \bibnamefont {Nowak}},\ and\ \bibinfo {author} {\bibfnamefont {I.}~\bibnamefont {Zahed}},\ }\bibfield  {title} {\bibinfo {title} {{Disorder in the Sachdev--Ye--Kitaev model}},\ }\href {https://doi.org/https://doi.org/10.1016/j.physletb.2017.08.054} {\bibfield  {journal} {\bibinfo  {journal} {Phys. Lett. B}\ }\textbf {\bibinfo {volume} {773}},\ \bibinfo {pages} {647} (\bibinfo {year} {2017})}\BibitemShut {NoStop}%
\bibitem [{\citenamefont {Gharibyan}\ \emph {et~al.}(2018)\citenamefont {Gharibyan}, \citenamefont {Hanada}, \citenamefont {Shenker},\ and\ \citenamefont {Tezuka}}]{Gharibyan18}%
  \BibitemOpen
  \bibfield  {author} {\bibinfo {author} {\bibfnamefont {H.}~\bibnamefont {Gharibyan}}, \bibinfo {author} {\bibfnamefont {M.}~\bibnamefont {Hanada}}, \bibinfo {author} {\bibfnamefont {S.~H.}\ \bibnamefont {Shenker}},\ and\ \bibinfo {author} {\bibfnamefont {M.}~\bibnamefont {Tezuka}},\ }\bibfield  {title} {\bibinfo {title} {Onset of random matrix behavior in scrambling systems},\ }\href {https://doi.org/10.1007/JHEP07(2018)124} {\bibfield  {journal} {\bibinfo  {journal} {JHEP}\ }\textbf {\bibinfo {volume} {2018}}\bibinfo  {number} { (7)},\ \bibinfo {pages} {124}}\BibitemShut {NoStop}%
\bibitem [{\citenamefont {Li}\ \emph {et~al.}(2021)\citenamefont {Li}, \citenamefont {Prosen},\ and\ \citenamefont {Chan}}]{DissipativeSFF}%
  \BibitemOpen
\bibfield  {number} {  }\bibfield  {author} {\bibinfo {author} {\bibfnamefont {J.}~\bibnamefont {Li}}, \bibinfo {author} {\bibfnamefont {T.}~\bibnamefont {Prosen}},\ and\ \bibinfo {author} {\bibfnamefont {A.}~\bibnamefont {Chan}},\ }\bibfield  {title} {\bibinfo {title} {Spectral statistics of non-hermitian matrices and dissipative quantum chaos},\ }\href {https://doi.org/10.1103/PhysRevLett.127.170602} {\bibfield  {journal} {\bibinfo  {journal} {Phys. Rev. Lett.}\ }\textbf {\bibinfo {volume} {127}},\ \bibinfo {pages} {170602} (\bibinfo {year} {2021})}\BibitemShut {NoStop}%
\bibitem [{\citenamefont {Matsoukas-Roubeas}\ \emph {et~al.}(2023)\citenamefont {Matsoukas-Roubeas}, \citenamefont {Beau}, \citenamefont {Santos},\ and\ \citenamefont {del Campo}}]{BrokenUnitaritySFF}%
  \BibitemOpen
  \bibfield  {author} {\bibinfo {author} {\bibfnamefont {A.~S.}\ \bibnamefont {Matsoukas-Roubeas}}, \bibinfo {author} {\bibfnamefont {M.}~\bibnamefont {Beau}}, \bibinfo {author} {\bibfnamefont {L.~F.}\ \bibnamefont {Santos}},\ and\ \bibinfo {author} {\bibfnamefont {A.}~\bibnamefont {del Campo}},\ }\bibfield  {title} {\bibinfo {title} {Unitarity breaking in self-averaging spectral form factors},\ }\href {https://doi.org/10.1103/PhysRevA.108.062201} {\bibfield  {journal} {\bibinfo  {journal} {Phys. Rev. A}\ }\textbf {\bibinfo {volume} {108}},\ \bibinfo {pages} {062201} (\bibinfo {year} {2023})}\BibitemShut {NoStop}%
\bibitem [{\citenamefont {Winer}\ \emph {et~al.}(2020)\citenamefont {Winer}, \citenamefont {Jian},\ and\ \citenamefont {Swingle}}]{Winer2020}%
  \BibitemOpen
  \bibfield  {author} {\bibinfo {author} {\bibfnamefont {M.}~\bibnamefont {Winer}}, \bibinfo {author} {\bibfnamefont {S.-K.}\ \bibnamefont {Jian}},\ and\ \bibinfo {author} {\bibfnamefont {B.}~\bibnamefont {Swingle}},\ }\bibfield  {title} {\bibinfo {title} {{Exponential Ramp in the Quadratic Sachdev-Ye-Kitaev Model}},\ }\href {https://doi.org/10.1103/PhysRevLett.125.250602} {\bibfield  {journal} {\bibinfo  {journal} {Phys. Rev. Lett.}\ }\textbf {\bibinfo {volume} {125}},\ \bibinfo {pages} {250602} (\bibinfo {year} {2020})}\BibitemShut {NoStop}%
\bibitem [{\citenamefont {Liao}\ \emph {et~al.}(2020)\citenamefont {Liao}, \citenamefont {Vikram},\ and\ \citenamefont {Galitski}}]{Liao2020}%
  \BibitemOpen
  \bibfield  {author} {\bibinfo {author} {\bibfnamefont {Y.}~\bibnamefont {Liao}}, \bibinfo {author} {\bibfnamefont {A.}~\bibnamefont {Vikram}},\ and\ \bibinfo {author} {\bibfnamefont {V.}~\bibnamefont {Galitski}},\ }\bibfield  {title} {\bibinfo {title} {{Many-Body Level Statistics of Single-Particle Quantum Chaos}},\ }\href {https://doi.org/10.1103/PhysRevLett.125.250601} {\bibfield  {journal} {\bibinfo  {journal} {Phys. Rev. Lett.}\ }\textbf {\bibinfo {volume} {125}},\ \bibinfo {pages} {250601} (\bibinfo {year} {2020})}\BibitemShut {NoStop}%
\bibitem [{\citenamefont {Berry}(1985)}]{BerrySpectralRigidity}%
  \BibitemOpen
  \bibfield  {author} {\bibinfo {author} {\bibfnamefont {M.~V.}\ \bibnamefont {Berry}},\ }\bibfield  {title} {\bibinfo {title} {Semiclassical theory of spectral rigidity},\ }\href {http://www.jstor.org/stable/2397963} {\bibfield  {journal} {\bibinfo  {journal} {Proceedings of the Royal Society of London. Series A, Mathematical and Physical Sciences}\ }\textbf {\bibinfo {volume} {400}},\ \bibinfo {pages} {229} (\bibinfo {year} {1985})}\BibitemShut {NoStop}%
\bibitem [{\citenamefont {{Ikeda}}\ \emph {et~al.}(2024)\citenamefont {{Ikeda}}, \citenamefont {{Vidmar}},\ and\ \citenamefont {{Flynn}}}]{OurPaper2}%
  \BibitemOpen
  \bibfield  {author} {\bibinfo {author} {\bibfnamefont {T.~N.}\ \bibnamefont {{Ikeda}}}, \bibinfo {author} {\bibfnamefont {L.}~\bibnamefont {{Vidmar}}},\ and\ \bibinfo {author} {\bibfnamefont {M.~O.}\ \bibnamefont {{Flynn}}},\ }\bibfield  {title} {\bibinfo {title} {{Exact spectral form factors of non-interacting fermions with Dyson statistics}},\ }\href {https://doi.org/10.48550/arXiv.2410.08269} {\bibfield  {journal} {\bibinfo  {journal} {arXiv e-prints}\ ,\ \bibinfo {eid} {arXiv:2410.08269}} (\bibinfo {year} {2024})},\ \Eprint {https://arxiv.org/abs/2410.08269} {arXiv:2410.08269 [cond-mat.stat-mech]} \BibitemShut {NoStop}%
\bibitem [{\citenamefont {Dyson}(1970)}]{DysonEigenvalueCorrelations}%
  \BibitemOpen
  \bibfield  {author} {\bibinfo {author} {\bibfnamefont {F.~J.}\ \bibnamefont {Dyson}},\ }\bibfield  {title} {\bibinfo {title} {{Correlations between the eigenvalues of a random matrix}},\ }\href {https://doi.org/10.1007/BF01646824} {\bibfield  {journal} {\bibinfo  {journal} {Commun. Math. Phys.}\ }\textbf {\bibinfo {volume} {19}},\ \bibinfo {pages} {235} (\bibinfo {year} {1970})}\BibitemShut {NoStop}%
\bibitem [{\citenamefont {Mehta}(1991)}]{mehta1991random}%
  \BibitemOpen
  \bibfield  {author} {\bibinfo {author} {\bibfnamefont {M.}~\bibnamefont {Mehta}},\ }\href {https://books.google.com/books?id=vBHSAQAACAAJ} {\emph {\bibinfo {title} {Random Matrices}}}\ (\bibinfo  {publisher} {Academic Press},\ \bibinfo {year} {1991})\BibitemShut {NoStop}%
\bibitem [{\citenamefont {Haake}(2006)}]{HaakeSignaturesofChaos}%
  \BibitemOpen
  \bibfield  {author} {\bibinfo {author} {\bibfnamefont {F.}~\bibnamefont {Haake}},\ }\href@noop {} {\emph {\bibinfo {title} {Quantum Signatures of Chaos}}}\ (\bibinfo  {publisher} {Springer-Verlag},\ \bibinfo {address} {Berlin, Heidelberg},\ \bibinfo {year} {2006})\BibitemShut {NoStop}%
\bibitem [{\citenamefont {Cipolloni}\ \emph {et~al.}(2023)\citenamefont {Cipolloni}, \citenamefont {Erd{\H{o}}s},\ and\ \citenamefont {Schr{\"o}der}}]{Cipolloni2023}%
  \BibitemOpen
  \bibfield  {author} {\bibinfo {author} {\bibfnamefont {G.}~\bibnamefont {Cipolloni}}, \bibinfo {author} {\bibfnamefont {L.}~\bibnamefont {Erd{\H{o}}s}},\ and\ \bibinfo {author} {\bibfnamefont {D.}~\bibnamefont {Schr{\"o}der}},\ }\bibfield  {title} {\bibinfo {title} {On the spectral form factor for random matrices},\ }\href {https://doi.org/10.1007/s00220-023-04692-y} {\bibfield  {journal} {\bibinfo  {journal} {Communications in Mathematical Physics}\ }\textbf {\bibinfo {volume} {401}},\ \bibinfo {pages} {1665} (\bibinfo {year} {2023})}\BibitemShut {NoStop}%
\bibitem [{Note1()}]{Note1}%
  \BibitemOpen
  \bibinfo {note} {For the purpose of numerical applications, the reader should note that eigenvalue solvers often report results in a prescribed order, e.g., extremal eigenvalues first. In the presence of interactions, permutation invariance of sites should be enforced by hand and we have done this by randomizing the quasienergy order following each unitary diagonalization.}\BibitemShut {Stop}%
\bibitem [{\citenamefont {Sierant}\ \emph {et~al.}(2023)\citenamefont {Sierant}, \citenamefont {Lewenstein}, \citenamefont {Scardicchio},\ and\ \citenamefont {Zakrzewski}}]{sierant_lewenstein_23}%
  \BibitemOpen
  \bibfield  {author} {\bibinfo {author} {\bibfnamefont {P.}~\bibnamefont {Sierant}}, \bibinfo {author} {\bibfnamefont {M.}~\bibnamefont {Lewenstein}}, \bibinfo {author} {\bibfnamefont {A.}~\bibnamefont {Scardicchio}},\ and\ \bibinfo {author} {\bibfnamefont {J.}~\bibnamefont {Zakrzewski}},\ }\bibfield  {title} {\bibinfo {title} {{Stability of many-body localization in Floquet systems}},\ }\href {https://doi.org/10.1103/PhysRevB.107.115132} {\bibfield  {journal} {\bibinfo  {journal} {Phys. Rev. B}\ }\textbf {\bibinfo {volume} {107}},\ \bibinfo {pages} {115132} (\bibinfo {year} {2023})}\BibitemShut {NoStop}%
\bibitem [{\citenamefont {Da{\u{g}}}\ \emph {et~al.}(2023)\citenamefont {Da{\u{g}}}, \citenamefont {Mistakidis}, \citenamefont {Chan},\ and\ \citenamefont {Sadeghpour}}]{Dag2023}%
  \BibitemOpen
  \bibfield  {author} {\bibinfo {author} {\bibfnamefont {C.~B.}\ \bibnamefont {Da{\u{g}}}}, \bibinfo {author} {\bibfnamefont {S.~I.}\ \bibnamefont {Mistakidis}}, \bibinfo {author} {\bibfnamefont {A.}~\bibnamefont {Chan}},\ and\ \bibinfo {author} {\bibfnamefont {H.~R.}\ \bibnamefont {Sadeghpour}},\ }\bibfield  {title} {\bibinfo {title} {Many-body quantum chaos in stroboscopically-driven cold atoms},\ }\href {https://doi.org/10.1038/s42005-023-01258-1} {\bibfield  {journal} {\bibinfo  {journal} {Communications Physics}\ }\textbf {\bibinfo {volume} {6}},\ \bibinfo {pages} {136} (\bibinfo {year} {2023})}\BibitemShut {NoStop}%
\end{thebibliography}%

\end{document}